\documentclass[a4paper,oneside,11pt]{article}

\usepackage{amsfonts,color}

\newtheorem{dfn}{Definition}[section]
\newtheorem{tw}[dfn]{Theorem}
\newtheorem{prop}[dfn]{Proposition}
\newtheorem{rem}[dfn]{Remark}

\newtheorem{lem}[dfn]{Lemma}

\newtheorem{cor}[dfn]{Corollary}
\usepackage{amssymb} 
\usepackage{amsmath}
\numberwithin{equation}{section}

\usepackage{anysize}
\usepackage{array}
\usepackage{enumerate}
\author{Micha\l \ Barski \\ \small  Faculty of Mathematics and Computer Science, University of Leipzig, Germany\\
\small Faculty of Mathematics, Cardinal Stefan Wyszy\'nski University in Warsaw, Poland
\\ \small{\it Michal.Barski@math.uni-leipzig.de}}

\title{\bf On the shortfall risk control\\ 
- a refinement of the quantile hedging method}

\begin{document}


\baselineskip=16pt
\maketitle
\date
\begin{abstract}

The issue of constructing a risk minimizing hedge under an additional almost-surely type constraint on the shortfall profile is examined.  
Several classical risk minimizing problems are adapted to the new setting and solved. In particular, the bankruptcy threat of optimal strategies 
appearing in the classical risk minimizing setting is ruled out.  The existence and concrete forms of optimal strategies in a general semimartingale market model with the use of conditional statistical tests are proven. The quantile hedging method applied in \cite{FL1} and \cite{FL2} as well as the classical Neyman-Pearson lemma are generalized. Optimal hedging strategies with shortfall constraints in the Black-Scholes and exponential  Poisson model are explicitly determined.
\end{abstract}

\noindent
\begin{quote}


\end{quote}


\section{Introduction}
Let us briefly sketch the classical hedging problem in a stochastic model of financial market. The goal of an investor having an initial capital $x\geq 0$ is to hedge dynamically a given random variable $H$ which represents the payoff of a financial contract at some future date $T>0$. He is looking for a trading strategy $\pi$ such that the related portfolio wealth $X^{x,\pi}_T$ at $T$ exceeds $H$ almost surely, i.e.
\begin{gather}\label{hedge wstep}
P(X^{x,\pi}_T\geq H)=1.
\end{gather}
A strategy $\pi$ satisfying \eqref{hedge wstep} is called a hedging or superhedging strategy for $H$ and it is well known that it exists if  $x$ is greater than the superhedging price of $H$. In the opposite case each trading strategy is able to hedge the claim at most partially, i.e. $P(X^{x,\pi}_T\geq H)<1$, and hence generates the shortfall 
$$
(H-X^{x,\pi}_T)^{+}:=\max\{0,H-X^{x,\pi}_T\}
$$ 
which is strictly positive with positive probability.  The related shortfall risk which appears in that case should be minimized to protect the investor against the resulting loss. There is an extensive literature on minimizing shortfall risk, see for instance \cite{CTW1}, \cite{CTW2}, \cite{Cvitanic}, \cite{FL1}, \cite{FL2}, \cite{KainaRuschendorf}, \cite{Melnikov1}, \cite{Melnikov2},\cite{Nakano}, \cite{Pham}, \cite{Rudloff}, \cite{Schied}, with various risk measures accepted by the investor. Four of those measures listed below play a central role in our study. In the quantile hedging approach, introduced in \cite{FL1}, the objective was  to maximize the probability of a successful hedge, i.e.
\begin{gather}\label{q.h. intro}
 \max_{\pi} P(X_T^{x,\pi}\geq H).
\end{gather}
The drawback of \eqref{q.h. intro} of neglecting the portfolio performance on the set $\{X_T^{x,\pi}< H\}$ has been partially eliminated in \cite{FL1} in the generalized version of \eqref{q.h. intro} which is
 \begin{gather}\label{g.q.h. intro}
 \max_{\pi}E\left[\varphi_{x,\pi}\right] ,\quad \text{where} \quad \varphi_{x,\pi}:=1\wedge \frac{X^{x,\pi}_T}{H}.
\end{gather}
Another optimality criterion was to minimize the weighted expected shortfall, i.e.
\begin{gather}\label{e.s. intro}
\min_{\pi} E[l((H-X_T^{x,\pi})^+)],
\end{gather}
where $l:\mathbb{R}\longrightarrow \mathbb{R}$ is a so called loss function.  The case $l(z)=z$ has been studied in \cite{Cvitanic} and the general case in \cite{FL2} and \cite{Pham}. If $l(z)=az+b, a\geq0, b\in\mathbb{R}$ 
then \eqref{e.s. intro} can be written as 
\begin{gather}\label{c.r.m. intro}
\min_\pi\rho\left(-(H-X^{x,\pi}_{T})^+\right),
\end{gather}
where $\rho$ is defined by $\rho(Y):=E[l(-Y)], Y\in L^1$. In this case $\rho$ is a coherent risk measure of a special form, see Section \ref{Formulations of the problems} for a precise definition. The general form of \eqref{c.r.m. intro} where $\rho: L^p\longrightarrow \mathbb{R}, p>1$ is a coherent risk measure on $L^p$ has been studied in \cite{Nakano}.

The motivation for the present paper arises from the fact that the profile of the shortfall in all the problems \eqref{q.h. intro}, \eqref{g.q.h. intro}, \eqref{e.s. intro} and \eqref{c.r.m. intro} remains beyond the trader's control. The preferences of the trader towards the size of the shortfall are not described sufficiently well by the risk measures mentioned above and consequently even a risk minimizing strategy may generate the loss which exceeds the solvency of the trader. So, even the best performance may lead to bankruptcy in finite time! This problem is apparent in the quantile hedging approach because, as shown in \cite{FL1}, the optimal strategy $\tilde{\pi}$ for \eqref{q.h. intro} is such that 
$$
X^{x,\tilde{\pi}}_T=H\mathbf{1}_{A},
$$ 
where $A$ is some subset of $\Omega$ which depends on $x$. It follows that the shortfall equals $H\mathbf{1}_{A^c}$ which means that the shortfall risk is completely unhedged on $A^c$. This depicts the quantile hedging method as a risky tool for minimizing the risk. 
Although the risk measures in \eqref{g.q.h. intro} and \eqref{e.s. intro} are more involved, the problem of an uncontrolled shortfall profile appears there as well. To illustrate that let us consider a call option $H=(S_T-K)^+$ on the underlying asset $S$ in the classical Black-Scholes model with drift $\alpha$ and volatility $\sigma>0$. It was shown in \cite[p. 261]{FL1}  that in the case when $\alpha<\sigma^2$ the optimal strategy $\tilde{\pi}$ for \eqref{g.q.h. intro} generates the wealth 
\begin{gather*}
X_T^{x,\tilde{\pi}}=(S_T-K)^{+}-(S_T-k)^+-(k-K)\mathbf{1}_{\{S_T>k\}},
\end{gather*}
where $k>K$ is a certain constant which depends on $x$. Thus the related shortfall equals  
\begin{gather*}
 H-X_T^{x,\tilde{\pi}}=(S_T-k)^++(k-K)\mathbf{1}_{\{S_T>k\}}.
\end{gather*}
In particular, it is clear that the shortfall is unbounded on the set $\{S_T>k\}$ which implies a
positive ruin probability for each investor regardless of the level of his solvency. An analogous example can be constructed for \eqref{e.s. intro} with $l(z)=z$ and the claim $H:=\frac{1}{S_T}$.

In this paper we show how to incorporate a relevant shortfall profile into the problems 
\eqref{q.h. intro}, \eqref{g.q.h. intro}, \eqref{e.s. intro} and \eqref{c.r.m. intro} which in turn allows to obviate the drawbacks of optimal strategies mentioned above. The idea is to introduce a shortfall constraint $L$ which is a random variable acting as upper bound for the shortfall and study the problems \eqref{q.h. intro}, \eqref{g.q.h. intro}, \eqref{e.s. intro}, \eqref{c.r.m. intro}   subject to the additional condition
\begin{gather}\label{shortfall restriction wstep}
P((H-X^{x,\pi}_{T})^+\leq L)=1.
\end{gather}
Since $L$ is of a fairly general form, this setting provides a flexible tool for managing hedging risk and allows to accommodate fully the risk preferences of the investor.  In particular, an appropriate choice of a bounded shortfall constraint protects him against a bankruptcy threat. Coming back to the example with a call option mentioned above, let us assume that the trader wants to keep the shortfall below a constant margin $c>0$. Our general results applied to this particular situation yield explicit  solutions to the problems \eqref{q.h. intro}, \eqref{g.q.h. intro} and \eqref{e.s. intro}. It turns out that the portfolio wealth of an optimal strategy is a digital combination of two options:
$(S_T-K)^{+}$ and $(S_T-(K+c))^+$.  A precise form of the combination depends on the risk measure and the initial capital $x$. The optimal strategy for \eqref{q.h. intro} satisfies
$$
X^{x,\tilde{\pi}}_T=(S_T-K)^{+}\mathbf{1}_{\{S_T\leq k_1\}\cup\{S_T\geq k_2\}}+(S_T-(K+c))^{+}\mathbf{1}_{\{k_1<S_T<k_2\}},
$$
with $K<k_1\leq K+c$ and $k_2\geq K+c$. For \eqref{g.q.h. intro} the optimal portfolio is such that 
$$
X^{x,\tilde{\pi}}_T=(S_T-K)^{+}\mathbf{1}_{\{S_T\leq k_3\}}+(S_T-(K+c))^{+}\mathbf{1}_{\{S_T>k_3\}},
$$
with $k_3>K$ while for \eqref{e.s. intro} with the loss function $l(z)=z$ we obtain
$$
X^{x,\tilde{\pi}}_T=(S_T-K)^{+}\mathbf{1}_{\{S_T> k_4\}}+(S_T-(K+c))^{+}\mathbf{1}_{\{S_T\leq k_4\}},
$$
with $k_4>K$. All the constants above depend on $x$.

In this paper we characterize optimal solutions for the problems \eqref{q.h. intro}, \eqref{g.q.h. intro}, \eqref{e.s. intro}, \eqref{c.r.m. intro} under \eqref{shortfall restriction wstep} for general forms of $H$ and $L$.  
For the sake of generality, \eqref{c.r.m. intro} will be analysed for a convex risk measure instead of a coherent one, see Section \ref{Formulations of the problems} for definition. Our assumptions concerning the market are rather weak because we require only that the price process $(S_t)$ is a locally bounded semimartingale and that the market is free of arbitrage in the sense that there is no free lunch with vanishing risk NFLVR. This setting enables us to use the dual characterization of the superhedging price proved in \cite{DS1}. Our investigation relies on a certain restriction of the success ratio $\varphi_{x,\pi}$, defined in \eqref{g.q.h. intro}, which is implied by \eqref{shortfall restriction wstep}. It allows to characterize the solutions of \eqref{q.h. intro}, \eqref{g.q.h. intro}, \eqref{e.s. intro}, \eqref{c.r.m. intro} in terms of certain statistical tests, i.e. $[0,1]$-valued random variables, which exceed a prespecified test $\varphi^\ast$. We call each test of this a conditional 
test with a rejection threshold $\varphi^{\ast}$. In the presented framework $\varphi^\ast$ is determined by $H$ and $L$. Our approach is a modification of the celebrated quantile hedging method applied in \cite{FL1} and \cite{FL2} and generalizes the results where the shortfall profile was unconstrained. This particular situation corresponds to the condition $L=H$ which generates the trivial rejection threshold $\varphi^\ast\equiv 0$. In Lemma \ref{lemat NP generalized} we prove a generalized version of the Neyman-Pearson lemma for conditional statistical tests. This enables us to find the explicit form of optimal solutions in the case when the market is complete and the laws of $ZH$ and $ZL$ are free of positive atoms, where $Z$ stands for the density of the martingale measure, see Proposition \ref{prop o postaci rozwiazan pomocniczych} for details. As a consequence we obtain a precise characterization of optimal strategies in the Black-Scholes and exponential Poisson model.

The paper is organized as follows. In Section \ref{Formulations of the problems} we describe the market model and formulate the optimization problems in a precise way. The main results characterizing optimal strategies with shortfall constraints are proven in Section \ref{Optimal strategies with shortfall constraints}. The concept of a conditional statistical test is discussed in Section  \ref{Generalized Neyman-Person lemma} where also a generalized version of the Neyman-Pearson lemma is proven and its application to determining optimal payoffs is shown. Examples are presented in Section \ref{Complete markets}.

\section{Formulations of the problems}\label{Formulations of the problems}

We will consider a continuous time financial market of a general form, studied in \cite{DS2}, where the stock price is given by an $\mathbb{R}^d$-valued locally bounded semimartingale $(S_t)_{t\in[0,T]}$ on a filtered probability space $(\Omega, (\mathcal{F}_t)_{t\in[0,T]},\mathcal{F}=\mathcal{F}_{T}, P)$ with $T<+\infty$. Trading positions are represented by pairs $(x, \pi)$, where $x\geq 0$ stands for an initial capital of the investor and $(\pi_t)_{t\in[0,T]}$ is a predictable, $S$-integrable stochastic process describing the trading strategy, that is, the wealth allocation between stocks. The resulting {\it gain process}
\begin{gather*}
(\pi\cdot S)_t:=\int_{0}^{t}\pi(s)dS(s), \qquad t\in[0,T],
\end{gather*}
is assumed to be uniformly bounded from below, i.e. there exists a constant $m\geq 0$ such that $(\pi\cdot S)_t\geq -m$ for each $t\in[0,T]$ a.s.. 
The class of all such strategies will be denoted by $\Pi$. If the {\it wealth process} corresponding to $(x,\pi)$, $x\geq0$, $\pi\in\Pi$  given by
\begin{gather*}
 X_t^{x,\pi}:=x+(\pi\cdot S)_t, \qquad t\in[0,T],
\end{gather*}
satisfies $ X_T^{x,\pi}\geq 0$, a.s.,  then  $(x,\pi)$ will be called {\it admissible}. For the sake of simplicity we  assume that the risk-free interest rate equals zero, i.e. $r=0$, so the value of $1$ Euro on a savings account is constant in time. Let us define
\begin{gather*}
K:=\{X=X^{0,\pi}_T=(\pi\cdot S)_T, \ \pi\in\Pi\},
\end{gather*}
a set of final portfolio values attained from zero initial capital and a set
\begin{gather*}
C:=\{Y\in L^\infty(\Omega,\mathcal{F}, P): Y\leq X \ \text {for some} \ X\in K\}
\end{gather*}
of all bounded random variables dominated by some element of $K$. 
The model admits {\it no free lunch with vanishing risk} NFLVR if
$$
\bar{C}\cap L^\infty_{+}(\Omega,\mathcal{F},P)=\{0\},
$$
where $\bar{C}$ stands for the closure of $C$ in $L^{\infty}(\Omega,\mathcal{F},P)$ and $ L^\infty_{+}(\Omega,\mathcal{F},P)$ consists of all non-negative elements of $L^{\infty}(\Omega,\mathcal{F},P)$. The NFLVR condition precludes arbitrage opportunities from the market, which means that there are no risk-free investments generating profits. If NFLVR is violated then there is a sequence of strategies $\pi^n\in\Pi$ such that  the corresponding gains satisfy
$$
(\pi^n \cdot S)_T\longrightarrow Y, \quad (\pi^n \cdot S)_T\geq Y-\frac{1}{n}, 
$$
for some $0\neq Y\in\bar{C}\cap L^\infty_{+}$. Hence $\{\pi^n\}$ realizes a positive profit at $T$ with zero initial cost and asymptotically vanishing risk because $(\pi^n \cdot S)_T\geq -\frac{1}{n}, n=1,2,...$ . Theorem 1 in \cite{DS1} provides a characterization of NFLVR which is
\begin{gather}\label{NFLVR eqivalence}
 NFLVR \quad \Longleftrightarrow \quad \mathcal{Q}\neq\emptyset,
\end{gather}
where $\mathcal{Q}$ stands for the set of all equivalent to $P$ probability measures under which $(S_t)$ is a 
local martingale. In this paper we will work under the NFLVR condition, so we make the standing assumption  $\mathcal{Q}\neq\emptyset$. 
A dual representation of superhedging prices arising from \eqref{NFLVR eqivalence}, which we describe below, plays
a central role in our analysis.

Let $H\geq0$ be an $\mathcal{F}_T$-measurable random variable representing a contingent claim with payoff at time $T$. An admissible strategy $(x,\pi), x\geq0, \pi\in\Pi$ is a {\it superhedge} of $H$ if
\begin{gather}\label{hedging}
P(X_T^{x,\pi}\geq H)=1.
\end{gather}
The {\it superhedging price} of $H$ is defined by
$$
p(H):=\inf\left\{x\geq0: \text{there exists} \ \pi\in\Pi \quad \text{such that} \quad P(X_T^{x,\pi}\geq H)=1\right\}.
$$
The dual characterization of $p(H)$ has been proven in \cite{DS1}, see Theorem 9 and Corollary 10.
\begin{tw}\label{tw DS1}
Under NFLVR the superhedging price admits the following dual representation
\begin{gather}\label{superhedging price}
 p(H)=\sup_{Q\in\mathcal{Q}}E^Q[H], 
\end{gather}
where $E^{Q}[\cdot]$ stands for the expectation under $Q$. Moreover, if $x=\sup_{Q\in\mathcal{Q}}E^Q[H]<+\infty$ then there exists $\pi\in\Pi$ such that $(x,\pi)$ satisfy \eqref{hedging}.  
\end{tw}
In the particular case when $\mathcal{Q}$ is a singleton the price of $H$ is given by the expectation of the claim under the unique martingale measure, i.e. $p(H)=E^Q[H]$. If the latter is finite then it follows from 
Theorem 16 in \cite{DS1} that the inequality in \eqref{hedging} holds as equality. Then the market is {\it complete} and a hedging strategy satisfying $X_T^{x,\pi}=H$ is called a {\it replicating strategy}. The concept of superhedging was introduced in \cite{El Karoui} where \eqref{superhedging price} was proven in the context of a concrete model driven by a multidimensional diffusion process. The equivalence \eqref{NFLVR eqivalence}
and formula \eqref{superhedging price} can be generalized to the case when $(S_t)$ is a semimartingale which is not necessarily locally bounded, see Theorem 1.1 and Theorem 5.12 in \cite{DS2}. In this case the set $\mathcal{Q}$, however, must be replaced with the set of all sigma-martingale measures and the superhedging strategy against $H$ 
for $x=\sup_{Q\in\mathcal{Q}}E^Q[H]<+\infty$ exists in an extended class of strategies satisfying certain technical conditions, see Theorem 5.5 in \cite{DS2}. The choice of a model with a locally bounded semimartingale $(S_t)$ seems to keep a good balance between the generality of our consideration and clarity of presentation.

Let us consider the situation when the initial capital $x$ does not allow to superhedge $H$, i.e. satisfies $0<x<p(H)$. It follows from Theorem \ref{tw DS1} that then \eqref{hedging} is violated and hence each strategy $\pi\in\Pi$ is biased by hedging risk with positive probability, that is
\begin{gather*}
 P((H-X_T^{x,\pi})^+>0)>0.
\end{gather*}
The problem of the investor is to minimize the risk which is quantified by a properly chosen risk measure. A component  of the risk measures considered in the sequel is a condition describing a maximal size of the shortfall $(H-X_T^{x,\pi})^+$ which is acceptable by the trader.
A {\it shortfall constraint} is defined by an $\mathcal{F}_T$-measurable random variable $L$ satisfying the condition 
\begin{gather}\label{shortfal constraint - ograniczenia}
0\leq L\leq H,
\end{gather}
and it constitutes those admissible strategies as acceptable which satisfy
\begin{gather}\label{war no bankruptcy}
P\Big((H-X_T^{x,\pi})^+\leq L\Big)=1.
\end{gather}
The condition \eqref{shortfal constraint - ograniczenia} precludes trading strategies with shortfall exceeding the value of the contract. It is intuitively clear that the initial capital and the shortfall constraint should be related to each other, that is portfolios with a restricted shortfall should keep the initial cost at a sufficiently high level. Indeed, due to the positivity of $L$, we have
$$
(H-X^{x,\pi}_{T})^+\leq L \quad \Longleftrightarrow \quad H-X^{x,\pi}_{T}\leq L,
$$
which implies that under \eqref{shortfal constraint - ograniczenia} the condition \eqref{war no bankruptcy} is equivalent to $P(X_T^{x,\pi}\geq H-L)=1$. Hence $(x,\pi)$ hedges the claim $H-L$ and consequently
\begin{gather}\label{x resrtiction}
 x\geq p(H-L).
\end{gather}
Below we give some natural examples of shortfall constraints corresponding to various forms of the trader's risk aversion.

\vskip2ex
\noindent
{\bf Examples}~
\begin{enumerate}[a)]
\item If $L=H$ then \eqref{war no bankruptcy} boils down to the positivity of $X^{x,\pi}_{T}$ and hence the profile of the shortfall remains unconstrained.  
This case corresponds to the classical framework considered in the literature. 
\item For $L=0$ the trader is expected to hedge the claim $H$, so no shortfall is acceptable at all.
\item The trader can cover the arising portfolio loss providing that it lies below a fixed constant level $c>0$. The maximal value of $c$ is defined by the solvency of the trader. In this case we set
\begin{gather*}
L= c\wedge H.
\end{gather*}
\item Generalizing the previous example, the trader may want to keep the loss below $c$ and simultaneously hedge $H$ in some fixed price range $[a,b]$ of the underlying stock. Then $L$ is given by
\begin{gather*}
 L=(c\wedge H)\mathbf{1}_{\{S_T<a\}}+(c\wedge H)\mathbf{1}_{\{S_T>b\}}.
\end{gather*}
\item In the subjective forecast of the trader the stock price range $(0,a)$, $(b,+\infty)$ is viewed as unrealistic and hence ruled out as source of risk.
The trader's aim is to keep the shortfall below $c$ only in the interval
$[a,b]$. The related form of $L$ is
\begin{gather*}
 L=(c\wedge H)\mathbf{1}_{\{S_T\in[a,b]\}}+H\mathbf{1}_{\{S_T\notin[a,b]\}}.
\end{gather*}
\item Let $\alpha\in[0,1]$ describe a partial recovery of the claim, i.e. the claim which is to be hedged is $\alpha H$. Then $L$ is equal to
\begin{gather*}
 L=(1-\alpha)H.
\end{gather*}
\end{enumerate}

Our aim is to solve the classical optimization problems \eqref{q.h. intro}, \eqref{g.q.h. intro}, \eqref{e.s. intro} and \eqref{c.r.m. intro} which are adapted to the new framework with a constrained shortfall profile. 
For a given triplet $(x,H,L)$, which in view of the discussion above satisfies $p(H)=\sup_{Q\in\mathcal{Q}}E^Q[H]<+\infty$, \eqref{shortfal constraint - ograniczenia}, \eqref{war no bankruptcy} and \eqref{x resrtiction}, we are looking for a strategy $\pi\in\Pi$ such that $(x,\pi)$ is admissible and solves
\begin{gather}\label{main problem}
\begin{cases}
&\min_{\pi\in\Pi}r(H,X^{x,\pi}_{T})\\[1ex]
\ (i) &P((H-X_T^{x,\pi})^+\leq L)=1\\[1ex]
\ (ii) &p(H-L)\leq x<p(H).
\end{cases}
\end{gather}
Recall that $p(H)$ stands for the price of $H$ and is given by \eqref{superhedging price}. Above $r(H,X^{x,\pi}_{T})$ describes the shortfall risk of the pair $(x,\pi)$ and \eqref{main problem} will be investigated for its four concrete forms. For  $$r(H,X^{x,\pi}_{T}):=P(X^{x,\pi}_{T}<H),$$
 we obtain the {\it quantile hedging problem} (QH). To the 
 {\it generalized quantile hedging problem} (GQH) corresponds
 $$r(H,X^{x,\pi}_{T}):=E\left[\left(1-\frac{X^{x,\pi}_{T}}{H}\right)\mathbf{1}_{\{X^{x,\pi}_{T}<H\}}\right],$$ 
 and to the {\it weighted expected shortfall problem} (WES)  
 $$r(H,X^{x,\pi}_{T}):=E[l((H-X^{x,\pi}_T)^+)].$$ 
The loss function in WES is state dependent, i.e. $l:\Omega\times [0,+\infty)\longrightarrow [0,+\infty)$ and such that $l(\omega,\cdot)$ is continuous, increasing and $l(\omega,0)=0$ for each $\omega\in\Omega$. We also assume that $E[l(H)]<+\infty$, so the risk measure of any admissible strategy in WES is well defined.
For the shortfall risk quantified by 
\begin{gather*}
 r\left(H,X^{x,\pi}_{T}\right)=\rho\left(-(H-X^{x,\pi}_{T})^+\right),
\end{gather*}
where $\rho$ is a convex risk measure, \eqref{main problem} will be denoted by CRM. 
Recall that 
$\rho:L^p\longrightarrow \mathbb{R}$ with $L^{p}=L^p(\Omega,\mathcal{F},P), p\geq 1$ is a convex risk measure if it satisfies
\begin{enumerate}[a)]
 \item $Z_1\leq Z_2 \quad \Longrightarrow \quad \rho(Z_1)\geq \rho(Z_2)$, \qquad $Z_1,Z_2\in L^p$,
 \item $\rho(Z+a)=\rho(Z)-a, \qquad a\in\mathbb{R}, Z\in L^p$,
 \item $\rho(\alpha Z_1+(1-\alpha)Z_2)\leq\alpha \rho(Z_1)+(1-\alpha)\rho(Z_2)$, \qquad $\alpha\in[0,1], Z_1,Z_2\in L^p$.
\end{enumerate}
If, additionally, 
\begin{enumerate}[d)]
 \item $\rho(\alpha Z)= \alpha\rho(Z)$, \qquad $\alpha\geq0, Z\in L^p$.
\end{enumerate}
is satisfied then $\rho$ is called coherent.

If $L=H$ then the constraints \eqref{main problem}(i) and \eqref{main problem}(ii) amount to the admissibility of $(x,\pi)$ and consequently \eqref{main problem} becomes a classical 
risk minimizing problem with an unconstrained shortfall profile.

\section{Optimal strategies with shortfall constraint}\label{Optimal strategies with shortfall constraints}

The following result describes an optimal strategy for the problem QH. Below $A^c$ stands for the complement of a set $A$.
\begin{prop}\label{prop pure quantile hedging}
Let $p(H-L)\leq x<p(H)$. If there exists a set $\tilde{A}\in\mathcal{F}$ solving the problem
\begin{gather}\label{problem pomocniczy pure quantile}
\begin{cases}
&\max_{A}P(A)\\[1ex]
\ (i) &p(H-L\mathbf{1}_{A^c})\leq x,
\end{cases}
\end{gather}
then a hedging strategy $(\tilde{x},\tilde{\pi})$ for the claim $\tilde{H}:=H-L\mathbf{1}_{\tilde{A}^c}$ with $\tilde{x}=p(\tilde{H})$ solves QH.
\end{prop}
{\bf Proof:} Let us define a success set of a strategy $(x,\pi)$ by
\begin{gather*}
 A_{x,\pi}:=\{X^{x,\pi}_{T}\geq H\}.
\end{gather*} 
First we show that for any strategy $(x,\pi)$ satisfying \eqref{main problem}$(i),(ii)$ we have
\begin{gather*}
 P(X^{x,\pi}_{T}\geq H)=P(A_{x,\pi})\leq P(\tilde{A}). 
\end{gather*}
Since $X_{T}^{x,\pi}\geq H$ on $A_{x,\pi}$ and, by \eqref{main problem}$(i)$, $X_{T}^{x,\pi}\geq H-L$ a.s., it follows 
\begin{gather*}
H-L\mathbf{1}_{A^c_{x,\pi}}=H\mathbf{1}_{A_{x,\pi}}+(H-L)\mathbf{1}_{A_{x,\pi}^c} \leq X_T^{x,\pi}.
\end{gather*}
Using the fact that $X^{x,\pi}$ is a $Q$-supermartingale for each $Q\in\mathcal{Q}$, we obtain
\begin{gather*}
 E^Q[H-L\mathbf{1}_{A^c_{x,\pi}}]\leq E^{Q}[X^{x,\pi}_T]\leq x,\quad Q\in\mathcal{Q},
\end{gather*}
which, by passing to supremum over $Q\in\mathcal{Q}$, gives
\begin{gather*}
 p(H-L\mathbf{1}_{A^c_{x,\pi}})\leq x,
\end{gather*}
and \eqref{problem pomocniczy pure quantile} $(i)$ follows. Hence $P(A_{x,\pi})\leq P(\tilde{A})$.

Now let us consider the strategy $(\tilde{x},\tilde{\pi})$ and notice that  the condition $X_T^{\tilde{x},\tilde{\pi}}\geq H-L\mathbf{1}_{\tilde{A}^c}$  implies 
\begin{gather}\label{poooooomocy}
X_T^{\tilde{x},\tilde{\pi}}\geq H\mathbf{1}_{\tilde{A}}+(H-L)\mathbf{1}_{\tilde{A}^c}\geq H-L,
\end{gather}
It follows that $(H-X_T^{\tilde{x},\tilde{\pi}})^+\leq L$, which is \eqref{main problem}$(i)$, and
that $\tilde{x}\geq p(H-L)$ which together with the condition $\tilde{x}=p(\tilde{H})\leq x$ gives
\eqref{main problem}$(ii)$. Further, it follows from \eqref{poooooomocy} that
$A_{\tilde{x},\tilde{\pi}}\supseteq \tilde{A}$
and thus $P(A_{\tilde{x},\tilde{\pi}})\geq P(\tilde{A})$. Hence $A_{\tilde{x},\tilde{\pi}}= \tilde{A}$ and the optimality of $(\tilde{x},\tilde{\pi})$ follows. \hfill$\square$

\vskip2mm
To deal with the succeeding problems we will need the success ratio of an admissible strategy $(x,\pi)$ defined by
\begin{gather}\label{success ratio}
 \varphi_{x,\pi}:=1\wedge\frac{X^{x,\pi}_T}{H},
\end{gather}
and a family of all statistical tests defined by
\begin{gather}\label{klasa R}
\mathcal{R}:=\{\varphi: \varphi \ \text{is} \ \mathcal{F}-\text{measurable and} \ 0\leq\varphi\leq 1\}.
\end{gather} 
It follows from \eqref{shortfal constraint - ograniczenia} follows that $\frac{H-L}{H}\in\mathcal{R}$ provided that $\frac{H-L}{H}$,
by definition, is equal to zero on the set $\{H=0\}$.
An optimal strategy for WES is characterized by the following result.

\begin{tw}\label{tw o optymalnej strategii (WES)}
Assume that the initial capital $x$ satisfies $p(H-L)\leq x<p(H)$. Let $\tilde{\varphi}\in\mathcal{R}$ be a solution of the problem
\begin{gather}\label{problem pomocniczy (WES)}
\begin{cases}
&\min_{\varphi}E[l((1-\varphi)H)]\\[1ex]
\ (i) &\varphi\geq\frac{H-L}{H},\\[1ex]
\ (ii) &p(H\varphi)\leq x.
\end{cases}
\end{gather}
Let $(\tilde{x},\tilde{\pi})$, $\tilde{x}=p(\tilde{H})$, be a hedging strategy for the claim $\tilde{H}:=H\tilde{\varphi}$. Then $(\tilde{x},\tilde{\pi})$ solves WES and $\varphi_{\tilde{x},\tilde{\pi}}=\tilde{\varphi}$.
\end{tw}
{\bf Proof:} Using the same type of arguments as in the proof of Proposition 3.1 in  \cite{FL2} we prove that $\tilde{\varphi}$ exists.
Let $\{\varphi\}_n$ be a minimizing sequence satisfying \eqref{problem pomocniczy (WES)}$(i), (ii)$.
There exists a new minimizing sequence
\begin{gather}\label{convex pomooocy}
\tilde{\varphi}_n\in conv\{\varphi_n,\varphi_{n+1},...\}
\end{gather}
which converges almost surely to a limit $\tilde{\varphi}$. Since $\tilde{\varphi}_n$ is an element of the convex hull \eqref{convex pomooocy} consisting of elements satisfying \eqref{problem pomocniczy (WES)}$(i), (ii)$ and since $p(\alpha H\varphi_k+\beta H\varphi_l)\leq \alpha p(H\varphi_k)+\beta p(H\varphi_l), l,k\geq 1, \alpha, \beta\geq0$ holds it follows that $\tilde{\varphi}_n$ also satisfies \eqref{problem pomocniczy (WES)}$(i), (ii)$. Hence $\tilde{\varphi}$ also satisfies \eqref{problem pomocniczy (WES)}$(i), (ii)$, so it solves \eqref{problem pomocniczy (WES)}.

Let $(x,\pi)$ be a strategy satisfying \eqref{main problem}$(i), (ii)$. It follows from \eqref{main problem}$(i)$ that
\begin{align*}
 \varphi_{x,\pi}&=1\wedge \frac{X^{x,\pi}_T}{H}\geq \mathbf{1}_{\{X^{x,\pi}_T\geq H\}}+\frac{H-L}{H}\mathbf{1}_{\{X^{x,\pi}_T< H\}}\\[1ex]
 &\geq \frac{H-L}{H}, 
\end{align*}
which implies that $\varphi_{x,\pi}$ satisfies \eqref{problem pomocniczy (WES)}$(i)$. Moreover, it follows from the inequality 
\begin{gather*}
 H\varphi_{x,\pi}=H\wedge X^{x,\pi}_T\leq X^{x,\pi}_T,
\end{gather*}
and from the fact that $X^{x,\pi}$ is a $Q$-supermartingale for each $Q\in\mathcal{Q}$ that
\begin{gather*}
p(H\varphi_{x,\pi})\leq p(X_T^{x,\pi})\leq x.
\end{gather*}
This  means that $\varphi_{x,\pi}$ satisfies \eqref{problem pomocniczy (WES)}$(ii)$. 
It follows that 
\begin{gather}\label{pierwsze szacowanie tw (WES)}
 E[l((H-X_T^{x,\pi})^+)]=E[l((1-\varphi_{x,\pi})H)]\geq E[l((1-\tilde{\varphi})H)].
\end{gather}
Now let us focus on the strategy $(\tilde{x}, \tilde{\pi})$. Since $\tilde{\varphi}$ satisfies \eqref{problem pomocniczy (WES)}$(i)$ it follows that 
\begin{gather}\label{popopopmocy}
X^{\tilde{x},\tilde{\pi}}_{T}\geq H\tilde{\varphi}\geq H\cdot \frac{H-L}{H}=H-L,
\end{gather}
and we obtain that $(H-X^{\tilde{x},\tilde{\pi}}_{T})^+\leq L$, which is \eqref{main problem}$(i)$. Furthermore, 
\eqref{popopopmocy} implies that $\tilde{x}\geq p(H-L)$, which together with the condition $\tilde{x}\leq x$ yields \eqref{main problem}$(ii)$. The success ratio of $(\tilde{x}, \tilde{\pi})$ satisfies
\begin{gather*}
 \varphi_{\tilde{x},\tilde{\pi}}=1\wedge \frac{X_T^{\tilde{x},\tilde{\pi}}}{H}\geq
 \mathbf{1}_{\{X^{\tilde{x},\tilde{\pi}}_T\geq H\}}+\tilde{\varphi}\mathbf{1}_{\{X^{\tilde{x},\tilde{\pi}}_T< H\}}\geq\tilde{\varphi},
\end{gather*}
and from the monotonicity of $l$ we obtain
\begin{gather}\label{drugie szacowanie tw (WES)}
 E[l((H-X_T^{\tilde{x},\tilde{\pi}})^+)] = E[l((1-\varphi_{\tilde{x},\tilde{\pi}})H)]\leq E[l((1-\tilde{\varphi})H)].
\end{gather}
The result follows from \eqref{pierwsze szacowanie tw (WES)} and \eqref{drugie szacowanie tw (WES)}.\hfill $\square$
\vskip2mm

The form of solution to GQH can be deduced from that of WES with the loss function
$$
\hat{l}(\omega, z):=\frac{z}{H(\omega)}\mathbf{1}_{\{H(\omega)>0\}}.
$$
Then, for any admissible strategy $(x,\pi)$, we have $E[\hat{l}((H-X^{x,\pi}_T)^+)]=E[\hat{l}((1-\varphi_{x,\pi})H)]=E[1-\varphi_{x,\pi}]$
and both problems GQH and WES are equivalent. This leads to the following result.

\begin{tw}\label{tw o optymalnej strategii (GQH)} 
Let $x$ be an arbitrary initial capital satisfying $p(H-L)\leq x<p(H)$. Denote by $\tilde{\varphi}\in\mathcal{R}$ a solution 
of the problem
\begin{gather}\label{problem pomocniczy}
\begin{cases}
&\max_{\varphi}E[\varphi]\\[1ex]
&\varphi\geq\frac{H-L}{H},\\[1ex]
&p(H\varphi)\leq x.
\end{cases}
\end{gather}
Then a hedging strategy $(\tilde{x},\tilde{\pi})$ with $\tilde{x}=p(\tilde{H})$ for the payoff $\tilde{H}:=H\tilde{\varphi}$ is optimal for the problem GQH and $\varphi_{\tilde{x},\tilde{\pi}}=\tilde{\varphi}$.
\end{tw}

The arguments in the poof of Theorem \ref{tw o optymalnej strategii (WES)} can be successfully applied in the case when the shortfall risk is given by
\begin{gather*}
 r\left(H,X^{x,\pi}_{T}\right)=\rho\left(-(H-X^{x,\pi}_{T})^+\right),
\end{gather*}
where $\rho$ stands for a convex risk measure on $L^p, p\geq 1$. Since each convex measure on $L^p$ is pointwise continuous, see Theorem 3.1 in \cite{KainaRuschendorf}, it follows that the pointwise convergent minimizing sequence $\{\varphi_n\}$ for the problem
$$
\min_{\varphi}\rho\left(-(1-\varphi)H\right), \quad \varphi\geq\frac{H-L}{H}, \quad p(H\varphi)\leq x,
$$
satisfies 
$$
\rho\left(-(1-\varphi_n)H\right)\underset{n}{\longrightarrow} \rho\left(-(1-\tilde{\varphi})H\right),
$$
where $\tilde{\varphi}:=\lim\varphi_n$. Following the proof of Theorem \ref{tw o optymalnej strategii (WES)} one can show that 
a hedging strategy for $H\tilde{\varphi}$ is optimal. Hence we obtain the following Corollary, which is a generalization of Theorem 1.5 in \cite{Nakano} dealing with minimizing coherent risk measures in the class of strategies with no shortfall constraints.

\begin{cor} Let $1\leq p<+\infty$. Assume that $E[H^p]<+\infty$ and $\rho$ is a convex risk measure on $L^p$.
For $x$ satisfying $p(H-L)\leq x<p(H)$ let $\tilde{\varphi}\in\mathcal{R}$ be a solution of the problem
\begin{gather}\label{problem pomocniczy convex measure}
\begin{cases}
&\min_{\varphi}\rho\left(-(1-\varphi)H\right)\\[1ex]
&\varphi\geq\frac{H-L}{H},\\[1ex]
&p(H\varphi)\leq x.
\end{cases}
\end{gather}
Let $(\tilde{x},\tilde{\pi})$, $\tilde{x}=p(\tilde{H})$, be a hedging strategy for the claim $\tilde{H}:=H\tilde{\varphi}$. Then $(\tilde{x},\tilde{\pi})$ solves CRM.
\end{cor}

We close this section with a remark on risk-independent hedging. 

\begin{rem} For a given shortfall constraint $L$ and $x$ such that $p(H-L)\leq x<p(H)$ the investor can be interested in finding a new shortfall constraints
$\tilde{L}\leq L$ and a feasible portfolio for $\tilde{L}$, i.e. satisfying $P((H-X^{x,\pi}_{T})^+\leq\tilde{L})=1$. Portfolios which are not feasible for $\tilde{L}$ are more risky than those feasible for $\tilde{L}$ regardless of the risk measure of the trader provided that it is monotone. Since the new constraint should preserve the profile of the original one, we can search it in the class  
 $\{L_\alpha:=\alpha L, \alpha\in[0,1]\}$. This leads to the problem
\begin{gather}\label{roboust hedging}
\begin{cases}
&\min\alpha\\[1ex]
&P((H-X^{x,\pi}_T)^+\leq L_{\alpha})=1,\\[1ex]
& p(H-L_\alpha)\leq x<p(H).
\end{cases}
\end{gather}
Since the function 
$$
g(\alpha):=p(H-L_{\alpha})=\sup_{Q\in\mathcal{Q}}E^Q[H-L_{\alpha}], \qquad \alpha\in[0,1],
$$ 
is continuous and monotone with $g(0)=p(H)$ and $g(1)=p(H-L)$, there exist solutions of the equation $g(\alpha)=x$ and $\hat{\alpha}:=\min\{\alpha\in[0,1]: g(\alpha)=x\}$. The solution of \eqref{roboust hedging} is $\hat{\alpha}$ and a feasible strategy for $L_{\hat{\alpha}}$
is a hedge for $H-\hat{\alpha}L$.

\end{rem}

\section{Generalized Neyman-Pearson lemma and complete markets}\label{Generalized Neyman-Person lemma}

In this section we analyse the conditions describing the success ratios of optimal strategies considered in Section \ref{Optimal strategies with shortfall constraints}. Notice that using \eqref{superhedging price}, which defines the superhedging price of $H$, the problem \eqref{problem pomocniczy} can be written in the form 
\begin{gather}\label{conditional statistical testing}
\begin{cases}
 \ (i)  &\max_{\varphi}E^P[\varphi] \\[1ex]
 \ (ii) &\varphi^\ast\leq \varphi\leq 1,\\[1ex]
 \ (iii) &\sup_{\hat{Q}\in\mathcal{\hat{Q}}} E^{\hat{Q}}[\varphi]\leq x.
\end{cases}
\end{gather}
where $\varphi^\ast:=\frac{H-L}{H}$ and $\mathcal{\hat{Q}}$ is the family of finite measures  defined by $d\hat{Q}:=H dQ$, $Q\in\mathcal{Q}$.
The conditions \eqref{conditional statistical testing} (i) and \eqref{conditional statistical testing}(iii) correspond to the classical problem of testing a null composite hypothesis represented by the family  $\hat{\mathcal{Q}}$ against a simple alternative hypothesis given by the measure $P$. More precisely, \eqref{conditional statistical testing}(iii) is a constraint for the type I statistical error while \eqref{conditional statistical testing}(i) describes minimization of the type II statistical error. The non-standard condition is \eqref{conditional statistical testing}(ii) which tells that each test must exceed the minimal threshold $\varphi^\ast$ of rejecting the null hypothesis. 
We call tests satisfying \eqref{conditional statistical testing}(ii), (iii) {\it conditional tests} with a {\it rejection threshold} $\varphi^\ast$. The rejection threshold affects of course both statistical errors. The error of the first kind is bounded from below, i.e.
$$
\sup_{\hat{Q}\in\mathcal{\hat{Q}}} E^{\hat{Q}}[\varphi] \geq \sup_{\hat{Q}\in\mathcal{\hat{Q}}} E^{\hat{Q}}[\varphi^\ast],
$$
while the error of the second kind is bounded from above, i.e.
$$
E^P[1-\varphi]\leq E^P[1-\varphi^\ast].
$$
It follows, in particular, that \eqref{conditional statistical testing} is well posed if $x\geq \sup_{\hat{Q}\in\mathcal{\hat{Q}}} E^{\hat{Q}}[\varphi^\ast]$. The special case when $\hat{\mathcal{Q}}$ is a singleton is of prime importance because it corresponds to complete markets which are analytically tractable. 
If this is the case and $\varphi^\ast=0$ then \eqref{conditional statistical testing} becomes a classical testing problem with simple hypotheses and its solution is described by the Neyman-Pearson lemma. There are several results in the literature which extend the classical 
Neyman-Pearson lemma to composite hypotheses, see  \cite{Cvitanic Karatzas}, \cite{Rudloff}, \cite{Rudloff Karatzas}, \cite{Schied}. The result proven below sets up a new kind of generalization concerned with conditional tests for simple hypotheses.

Recall from \eqref{klasa R}, that $\mathcal{R}$ stands for the family of statistical tests.

\begin{lem}\label{lemat NP generalized} Let $P$ and $Q$ be any two equivalent probability measures.
For given $\varphi^\ast\in\mathcal{R}$ and $\alpha\in[E^Q[\varphi^\ast],1]$ a solution $\tilde{\varphi}$ of the problem

\begin{gather}\label{NP lemma - generalized}
\begin{cases}
&\max_{\varphi}E^{P}[\varphi]\\[1ex]
\ (i) &\varphi^\ast\leq\varphi\leq 1,\\[1ex]
\ (ii) &E^{Q}[\varphi]\leq \alpha,
\end{cases}
\end{gather}
has the form
\begin{gather}\label{NP-postac rozwiazania}
 \tilde{\varphi}=\mathbf{1}_{\{\varphi^\ast=1\}\cup\{\frac{dP}{dQ}>k\}}+[\varphi^\ast+\gamma(1-\varphi^\ast)]\mathbf{1}_{\{\frac{dP}{dQ}=k\}}+\varphi^\ast\mathbf{1}_{\{\frac{dP}{dQ}<k\}},
\end{gather}
where $k\geq0,\gamma\in[0,1]$ are constants such that $E^{Q}[\tilde{\varphi}]=\alpha$.
\end{lem}
{\bf Proof:} It is clear that $\tilde{\varphi}=\varphi^\ast=1$ on the set $\{\varphi^\ast=1\}$. On the set $\{\varphi^\ast<1\}$ the optimal solution $\tilde{\varphi}$ solves the problem
\begin{gather}\label{NP-2}
\begin{cases}
&\max_{\varphi}E^{P}[\varphi\mathbf{1}_{\{\varphi^\ast<1\}}]\\[1ex]
\ (i) &\varphi^\ast\mathbf{1}_{\{\varphi^\ast<1\}}\leq\varphi\mathbf{1}_{\{\varphi^\ast<1\}}\leq 1,\\[1ex]
\ (ii) &E^{Q}[\varphi\mathbf{1}_{\{\varphi^\ast<1\}}]\leq \alpha-P(\varphi^\ast=1).
\end{cases}
\end{gather}
For any $\varphi$ such that $\varphi^\ast\leq\varphi\leq1$ consider the transformation 
\begin{gather}\label{wzor na transformacje}
\Phi=\Phi(\varphi):=\frac{\varphi\mathbf{1}_{\{\varphi^\ast<1\}}-\varphi^\ast\mathbf{1}_{\{\varphi^\ast<1\}}}{(1-\varphi^\ast)\mathbf{1}_{\{\varphi^\ast<1\}}},
\end{gather}
which defines a random variable on the set $\hat{\Omega}:=\{\varphi^\ast<1\}$. 
The problem \eqref{NP-2} can be transformed with the use of two auxiliary probability measures on $\hat{\Omega}$ with densities
$$
\frac{d\hat{P}}{dP}:=
\frac{(1-\varphi^\ast)\mathbf{1}_{\{\varphi^\ast<1\}}}{E^P[(1-\varphi^\ast)\mathbf{1}_{\{\varphi^\ast<1\}}]}, \qquad
\frac{d\hat{Q}}{dQ}:=\frac{(1-\varphi^\ast)\mathbf{1}_{\{\varphi^\ast<1\}}}{E^Q[(1-\varphi^\ast)\mathbf{1}_{\{\varphi^\ast<1\}}]},
$$
to the form
\begin{gather}\label{NP-3}
\begin{cases}
&\max_{\Phi}E^{\hat{P}}[\Phi]\\[1ex]
\ (i) &0\leq \Phi \leq 1,\\[1ex]
\ (ii) &E^{\hat{Q}}[\Phi]\leq \frac{\alpha-Q(\varphi^\ast=1)-E^Q[(1-\varphi^\ast)\mathbf{1}_{\{\varphi^\ast<1\}}]}{E^Q[(1-\varphi^\ast)\mathbf{1}_{\{\varphi^\ast<1\}}]}.
\end{cases}
\end{gather}
The problem \eqref{NP-3} is a standard testing problem and the classical Neyman-Pearson lemma provides its solution
\begin{gather}\label{Phi 1}
\tilde{\Phi}=\mathbf{1}_{\{\frac{d\hat{P}}{d\hat{Q}}>k\}}+\gamma\mathbf{1}_{\{\frac{d\hat{P}}{d\hat{Q}}=k\}},
\end{gather}
where $k\geq0, \gamma\in[0,1]$ are constants such that \eqref{NP-3} (ii) holds as equality.
Since 
$$
\frac{d\hat{P}}{d\hat{Q}}=const. \frac{dP}{dQ}\mathbf{1}_{\{\varphi^\ast<1\}}, \quad const.>0,
$$
the optimal solution of \eqref{NP-3} can be written in the form
\begin{gather}\label{Phi 2}
\tilde{\Phi}=\mathbf{1}_{\{\frac{dP}{dQ}\mathbf{1}_{\{\varphi^\ast<1\}}>k\}}+\gamma\mathbf{1}_{\{\frac{dP}{dQ}\mathbf{1}_{\{\varphi^\ast<1\}}=k\}},
\end{gather}
where the constant $k$ in \eqref{Phi 1} and \eqref{Phi 2} may differ. Coming back to \eqref{wzor na transformacje} we determine  $\tilde{\varphi}\mathbf{1}_{\{\varphi^\ast<1\}}$ from the equation
$$
\tilde{\Phi}=\tilde{\Phi}(\tilde{\varphi})=\mathbf{1}_{\{\frac{dP}{dQ}\mathbf{1}_{\{\varphi^\ast<1\}}>k\}}+\gamma\mathbf{1}_{\{\frac{dP}{dQ}\mathbf{1}_{\{\varphi^\ast<1\}}=k\}},
$$
which gives 
$$
\tilde{\varphi}\mathbf{1}_{\{\varphi^\ast<1\}}=\mathbf{1}_{\{\frac{dP}{dQ}>k\}}
+[\varphi^\ast+\gamma(1-\varphi^\ast)]\mathbf{1}_{\{\frac{dP}{dQ}=k\}}+\varphi^\ast\mathbf{1}_{\{\frac{dP}{dQ}<k\}}.
$$
This, in view of the decomposition $\tilde{\varphi}=\tilde{\varphi}\mathbf{1}_{\{\varphi^\ast=1\}}+\tilde{\varphi}\mathbf{1}_{\{\varphi^\ast<1\}}$,
yields \eqref{NP-postac rozwiazania}.
\hfill$\square$
\vskip2ex
\noindent
One can check that Lemma \ref{lemat NP generalized} with $\varphi^\ast=0$ boils down to the classical Neyman-Pearson lemma.
\vskip2mm
The following part of this section is concerned with a precise characterization of solutions to problems 
\eqref{problem pomocniczy pure quantile}, \eqref{problem pomocniczy (WES)}, \eqref{problem pomocniczy}, \eqref{roboust hedging} in the case when the market is complete. Let us start with an auxiliary technical result.

\begin{prop}\label{prop pomocniczy o ciaglosci}
 Let $X\geq 0$, $Y\geq0$ be random variables with $E[Y]<+\infty$ and such that the cumulative distribution function
 $$
 F_X(t):=P(X\leq t), \quad t>0,
 $$
 is continuous. Then the function
 $$
 g(k):=E[Y\mathbf{1}_{\{X<k\}}], \quad k>0,
 $$
 is continuous.
 \end{prop}
{\bf Proof:} If $Y=0$ then the result is obvious. For $Y\neq 0$ let us consider the measure $\hat{P}$ given by $\frac{d\hat{P}}{dP}=\frac{Y}{E[Y]}$, 
which is clearly absolutely continuous with respect to $P$.  For $k>0$ we have
\begin{gather*}
 \left\vert g(k+\frac{1}{n})-g(k))\right\vert=E[Y\mathbf{1}_{\{k\leq X<k+\frac{1}{n}\}}]
 =E[Y]\hat{P}\left(k\leq X<k+\frac{1}{n}\right)
 \underset{n\rightarrow +\infty}{\longrightarrow} E[Y]\hat{P}(X=k)=0,
 \end{gather*}
 where the last equality follows from the absolute continuity of $\hat{P}$ with respect to $P$ and the continuity of $F_X$. The left continuity of $g$ follows from
\begin{gather*}
 \mid g(k)-g(k-\frac{1}{n})\mid=E[Y\mathbf{1}_{\{k-\frac{1}{n}\leq X<k\}}]\underset{n\rightarrow +\infty}{\longrightarrow} 0, \quad k>0,
\end{gather*}
which is a consequence of the dominated convergence.\hfill $\square$

\vskip2mm

Now we are ready to formulate conditions which ensure the existence of solutions to problems  \eqref{problem pomocniczy pure quantile}, \eqref{problem pomocniczy (WES)}, \eqref{problem pomocniczy}, \eqref{roboust hedging} and 
give their explicit form.

\begin{prop}\label{prop o postaci rozwiazan pomocniczych}
 Assume that the market model is complete and denote by $Z:=\frac{dQ}{dP}>0$ the density of the unique equivalent martingale measure $Q$.
 \begin{enumerate}[a)]
  \item Assume that the function $ F_{ZL}(t)=P(ZL\leq t), \ t>0,$
is continuous. Then \eqref{problem pomocniczy pure quantile} has a solution $\tilde{A}$ of the form
\begin{align}\label{postac Atilde w proposition}
\tilde{A}&=\{ZL<k\}, \quad \text{if}  \quad E^Q[H-L]<x<E^Q[H],\\[1ex]\nonumber
\tilde{A}&=\emptyset, \quad \text{if} \quad E^Q[H-L]=x,
\end{align}
where $k$ is a constant solving the equation $E[ZL\mathbf{1}_{\{ZL<k\}}]=x-E^Q[H-L]$.
\item Assume that the function $F_{ZH}(t)=P(ZH\leq t), \ t>0,$ is continuous. Then the solution of \eqref{problem pomocniczy} is given by
\begin{align}\label{postac tildefi w dowodzie}\nonumber
\tilde{\varphi}&=\frac{H}{H-L} \quad \text{if} \quad E^Q[H-L]=x,\\[1ex]
\tilde{\varphi}&=\mathbf{1}_{\{L=0\}\cup\{ZH<k\}}+\frac{H-L}{L}\mathbf{1}_{\{ZH>k\}}
\quad \text{if}  \quad E^Q[H-L]<x<E^Q[H],
\end{align}
with the constant $k$ solving
\begin{gather}\label{warunek na k GQH w dowodzie}
 E[ZL \mathbf{1}_{\{ZH>k\}}]=E^Q[H]-x.
\end{gather}
\item If the function $F_Z(t)=P(Z\leq t), t\in\mathbb{R}$, is continuous then the solution of \eqref{problem pomocniczy (WES)} with $l(z)=z$ equals
\begin{align*}
\tilde{\varphi}&=\frac{H}{H-L} \quad \text{if} \quad E^Q[H-L]=x,\\[1ex]
\tilde{\varphi}&=\mathbf{1}_{\{L=0\}\cup\{Z<k\}}+\frac{H-L}{L}\mathbf{1}_{\{Z>k\}}
\quad \text{if}  \quad E^Q[H-L]<x<E^Q[H].
\end{align*}
\item The solution of \eqref{roboust hedging} is equal to
$$
L_{\tilde{\alpha}}=\frac{E^Q[H]-x}{E^Q[L]} \ L.
$$
 \end{enumerate}
\end{prop}
{\bf Proof:} $a)$ By rearranging terms and introducing the auxiliary measure
$\hat{Q}$ defined by $d\hat{Q}:=\frac{L}{E^Q[L]}dQ$ one can reformulate
\eqref{problem pomocniczy pure quantile} to the form
\begin{gather}\label{problem pomocniczy pure quantile przerobiony}
\begin{cases}
&\max_{A} P(A)\\[1ex]
\ (i) &{\hat Q}[A]\leq \frac{x-E^Q[H-L]}{E^Q[L]}.
\end{cases}
\end{gather}
If $x=E^{Q}[H-L]$ then the solution of \eqref{problem pomocniczy pure quantile przerobiony} is $\tilde{A}=\emptyset$. If $E^{Q}[H-L]<x<E^Q[H]$ then  
$0<\frac{x-E^Q[H-L]}{E^Q[L]}<1$ and from the classical Neyman-Pearson lemma it follows that the solution of \eqref{problem pomocniczy pure quantile przerobiony} has the form
$$
\tilde{A}=\left\{\frac{dP}{d\hat{Q}}>c\right\}=\left\{ZL<\frac{E^Q[L]}{c}\right\},
$$ 
providing that the constant $c>0$ solves the equation
\begin{gather}\label{warunek na stala k w dowodzie}
\hat{Q}(\tilde{A})=\frac{1}{E^Q[L]}E\left[ZL \ \mathbf{1}_{\{ZL<\frac{E^Q[L]}{c}\}}\right]=\frac{x-E^Q[H-L]}{E^Q[L]}.
\end{gather}
Now we argue that \eqref{warunek na stala k w dowodzie} actually has a solution.
In view of Proposition \ref{prop pomocniczy o ciaglosci} the function
$$
g(c):=\frac{1}{E^Q[L]}E\left[ZL \ \mathbf{1}_{\{ZL<\frac{E^Q[L]}{c}\}}\right],
$$
is continuous on $(0,+\infty)$. Since $\lim_{c\rightarrow 0}g(c)=1$ and 
$\lim_{c\rightarrow +\infty}g(c)=0$ the existence of a solution to $g(c)=\frac{x-E^Q[H-L]}{E^Q[L]}$
follows. Equation \eqref{postac Atilde w proposition} holds with $k:=\frac{E^Q[L]}{c}$. \\

\noindent
$b)$  For $x=E^Q[H-L]$ we obtain immediately that $\tilde{\varphi}=\frac{H-L}{H}$ solves \eqref{tw o optymalnej strategii (GQH)}. Let us consider the case with $E^Q[H-L]<x<E^Q[H]$ and reformulate \eqref{tw o optymalnej strategii (GQH)} to the form required in Lemma \ref{lemat NP generalized}, that is
\begin{gather*}
\begin{cases}
&\max{\varphi} E[\varphi]\\[1ex]
\ (i) &\varphi\geq \frac{H-L}{H},\\[1ex]
\ (ii) &E^{\bar{Q}}[\varphi]\leq \frac{x}{E^Q[H]},
\end{cases}
\end{gather*}
with $\frac{d\bar{Q}}{dQ}:=\frac{H}{E^Q[H]}$. Since $F_{ZH}(\cdot)$ is continuous and hence $\frac{dP}{d\bar{Q}}=\frac{E^Q[H]}{ZH}$ atom-free distributed, it follows that the form of solution \eqref{NP-postac rozwiazania} given by Lemma \ref{lemat NP generalized} can be reduced to
\begin{align*}
 \tilde{\varphi}&=\mathbf{1}_{\{\frac{H-L}{H}=1\}\cup\{\frac{dP}{d\bar{Q}}>c\}}+\frac{H-L}{H}\mathbf{1}_{\{\frac{dP}{d\bar{Q}}<c\}}\\[1ex]
 &=\mathbf{1}_{\{L=0\}\cup\{ZH<\frac{E^Q[H]}{c}\}}+\frac{H-L}{H}\mathbf{1}_{\{ZH>\frac{E^Q[H]}{c}\}},
\end{align*}
with the constant $c$ solving $E^{\bar{Q}}[\tilde{\varphi}]=\frac{x}{E^Q[H]}$. The latter equation can be written in the form 
$$
E[ZL\mathbf{1}_{\{ZH>\frac{E^Q[H]}{c}\}}]=E^Q[H]-x, 
$$
which is \eqref{warunek na k GQH w dowodzie}. 
Notice that \eqref{warunek na k GQH w dowodzie} has a solution because $0<E^Q[H]-x< E^Q[L]$ holds, the function 
$$
h(c):=E[ZL\mathbf{1}_{\{ZH>\frac{E^Q[H]}{c}\}}], \quad c>0,
$$
is continuous by Proposition \ref{prop pomocniczy o ciaglosci} and
satisfies
\begin{gather*}
 \lim_{c\rightarrow 0}h(c)=0,
 \lim_{c\rightarrow +\infty}h(c)=E[ZL\mathbf{1}_{\{H>0\}}]=E[ZL]-E[ZL\mathbf{1}_{\{H=0\}}]=E[ZL]=E^Q[L].
\end{gather*}
Setting $k:=\frac{E^Q[H]}{c}$ we obtain the assertion.

\noindent
$c)$ With two auxiliary measures $\bar{P}, \bar{Q}$ given by $\frac{d\bar{P}}{dP}=\frac{H}{E[H]}$ and $\frac{d\bar{Q}}{dQ}=\frac{H}{E^Q[H]}$ one can mimic the proof of $(b)$.

\noindent
$d)$ Follows immediately because $\tilde{\alpha}$ solves the equation $E^Q[H-\tilde{\alpha} L]=x$. \hfill$\square$

\section{Concrete complete markets}\label{Complete markets}
 Our aim now is to minimize the hedging risk of a call option $(S_T-K)^+, K>0$ in the class of strategies subject to the shortfall constraint
 $L=c\wedge(S_T-K)^+$ with $c\geq 0$. The Black-Scholes  and exponential Poisson models will be examined.
 The initial capital of the investor is assumed to satisfy \eqref{x resrtiction}, which amounts to
$$
p\Big((S_T-(K+c))^+\Big)\leq x <p\Big((S_T-K)^+\Big).
$$
This means that $x$ is less than the replicating cost of the option but is also greater than the replicating cost of the call option with strike $K+c$. Combining Proposition \ref{prop pure quantile hedging}, Theorem \ref{tw o optymalnej strategii (GQH)}, Theorem \ref{tw o optymalnej strategii (WES)} together with Lemma \ref{lemat NP generalized} and Proposition \ref{prop o postaci rozwiazan pomocniczych} we show in the following paragraphs that an optimal strategy hedges always an option which is a sum of two knock-out options, i.e. it has the form
\begin{gather}\label{postac rozwiazania}
\tilde{H}=(S_T-K)^+\mathbf{1}_{A}+(S_T-(K+c))^+\mathbf{1}_{A^c},
\end{gather}
where $A\in\mathcal{F}_T$ is a set which depends on the initial capital $x$ and the risk measure of the investor. For the exponential Poisson model an additional term in \eqref{postac rozwiazania} appears which is related to the presence of jumps of the price process, see formula \eqref{exp Poisson - optimal payoff} below.

\subsection{Black-Scholes model}
 Let us recall some basics concerning the Black-Scholes model. The asset price dynamics has the form
\begin{gather*}
dS_t=S_t(\alpha dt+\sigma dW_t), \ S_0=s_0, \quad t\in[0,T], \quad \alpha\in\mathbb{R}, \sigma>0.
\end{gather*}
The unique martingale measure $Q$ is given by
\begin{gather*}
 \frac{dQ}{dP}=Z=e^{-\theta W_T-\frac{1}{2}\theta^2T}, 
\end{gather*}
with $\theta=\frac{\alpha}{\sigma}$. Under $Q$ the process $\tilde{W}_t:=W_t+\theta t$ is a Wiener process and the dynamics of $S$ under $Q$ has the form $dS_t=\sigma d\tilde{W}_t$.  The price of the call option is given by
$$
C_{BS}(K):=p\big((S_T-K)^{+}\big)=E^Q[(S_T-K)^{+}]=s_0\phi(d_1)-K\phi(d_2),
$$
where 
\begin{gather*}
 d_1:=\frac{\ln\left(\frac{s_0}{K}\right)+\frac{1}{2}\sigma^2T}{\sigma\sqrt{T}}, \qquad d_2:=d_1-\sigma\sqrt{T},
\end{gather*} 
and $\Phi$ stands for the $N(0,1)$-cumulative distribution function. 

Below we solve the problems QH, GQH and WES explicitly. First notice that if $x=C_{BS}(K+c)$ then QH, GQH and WES have the same solution which is the replicating strategy for the payoff $(S_T-(K+c))^+$. Hence in the sequel we consider the case $C_{BS}(K+c)<x<c_{BS}(K)$. For the sake of simplicity the parameters are assumed to satisfy $0<\alpha<\sigma^2$. Another cases can be treated, however, in a similar way. 

\vskip2mm
\noindent
{\bf Quantile hedging problem (QH)}
\begin{prop} \label{prop Black-Scholes pure q-h}
Let $C_{BS}(K+c) <x<C_{BS}(K)$. An optimal strategy for a call option $(S_T-K)^+$ with the shortfall constraint $L=c\wedge (S_T-K)^+$ in the Black-Scholes model with parameters satisfying  $0<\alpha<\sigma^2$ is a replicating strategy for the payoff
 \begin{gather}\label{optymalna wyplata dla call}
  \tilde{H}=(S_T-K)^+\mathbf{1}_{\{S_T\leq I(k)\}}+(S_T-K)^+\mathbf{1}_{\{S_T\geq J(k)\}}+(S_T-(K+c))^+\mathbf{1}_{\{I(k) < S_T < J(k)\}},
 \end{gather}
where 
\begin{align*}
 I(k)&:=\hat{y}(k)\wedge (K+c),\\[1ex]
 J(k)&:= (Clk)^{\frac{\sigma^2}{\alpha}}\vee (K+c),
\end{align*}
with $C:=\Big(\frac{1}{s_0}e^{-\frac{1}{2}(\alpha+\sigma^2)T}\Big)^{-\frac{\alpha}{\sigma^2}}$ and $\hat{y}(k)$ being the unique solution of the equation
\begin{gather*}
\frac{1}{Ck}y^{\frac{\alpha}{\sigma^2}}=y-K, \quad y\geq 0,
\end{gather*}
The constant $k$ in \eqref{optymalna wyplata dla call} is uniquely defined by the relation
\begin{gather}\label{warunek na c}
C_{BS}(K)+C_{BS}(K+c)-C_{BS}(I(k))-(I(k)-K)(1-\Phi(e_1(k)))+c(1-\Phi(e_2(k)))=x,
\end{gather}
where
\begin{gather*}
 e_1(k):=\frac{\ln\left(\frac{I(k)}{s_0}\right)+\frac{1}{2}\sigma^2T}{\sigma\sqrt{T}}, \qquad e_2(k):=\frac{\ln\left(\frac{J(k)}{s_0}\right)+\frac{1}{2}\sigma^2T}{\sigma\sqrt{T}}.
\end{gather*}
In particular, the shortfall of the optimal strategy equals $[S_T\wedge (K+c)-K]\mathbf{1}_{\{I(k)< S_T< J(k)\}}$.
\end{prop}
{\bf Proof:} Since $Z=CS_T^{-\frac{\alpha}{\sigma^2}}$ and $0<\alpha<\sigma^2$ it follows from Proposition \ref{prop o postaci rozwiazan pomocniczych} $(a)$ that the form of a solution to \eqref{problem pomocniczy pure quantile} is
\begin{align*}
\tilde{A}&=\{ZL<\frac{1}{k}\}=\{S_T\geq(kcC)^{\frac{\sigma^2}{\alpha}}\vee (K+c)\}\cup\{S_T\leq \hat{y}(k)\wedge (K+c)\}\\[1ex]
&=\{S_T\leq I(k)\}\cup\{S_T\geq J(k)\}.
\end{align*}
Hence the optimal payoff given by Proposition \ref{prop pure quantile hedging} is
\begin{align}\label{wzor pomocniczy}\nonumber
\tilde{H}&=\tilde{H}(k)=(S_T-K)^+\mathbf{1}_{\tilde{A}}+\Big((S_T-K)^+-(S_T-K)^+\wedge c\Big)\mathbf{1}_{\tilde{A}^c}\\[1ex]
&=(S_T-K)^+\mathbf{1}_{\{S_T\leq I(k)\}}+(S_T-K)^+\mathbf{1}_{\{S_T\geq J(k)\}}+(S_T-(K+c))^+\mathbf{1}_{\{I(k) < S_T < J(k)\}}.
\end{align}
To get an explicit characterization of $k$ let us decompose 
$\tilde{H}$ into the form
\begin{align*}
 \tilde{H} &=(S_T-K)^++(S_T-(K+c))^+-(S_T-I(k))^+
  -(I(k)-K)\mathbf{1}_{\{S_T>I(k)\}}+c\mathbf{1}_{\{S_T>J(k)\}},
\end{align*}
which allows to determine the price of $\tilde{H}$ in terms of the Black-Scholes call prices. Hence the equation for $k$ is 
$$
C_{BS}(K)+C_{BS}(K+c)-C_{BS}(I(k))-(I(k)-K)Q(S_T>I(k))+cQ(S_T\geq J(k))=x
$$
which leads directly to \eqref{warunek na c}. \hfill$\square$

\vskip2mm
\noindent
{\bf Generalized quantile hedging problem (GQH)}
\begin{prop}\label{prop Black-Scholes generalized g-h}
Let $C_{BS}(K+c)< x<C_{BS}(K)$. An optimal strategy for a call option $(S_T-K)^+$ with the shortfall constraint $L=c\wedge (S_T-K)^+$ in the Black-Scholes model with parameters satisfying  $0<\alpha<\sigma^2$ is a replicating strategy for the payoff
 \begin{gather*}
  \tilde{H}=(S_T-K)^{+}\mathbf{1}_{\{S_T\leq \hat{y}(k)\}}+(S_T-(K+c))^{+}\mathbf{1}_{\{S_T>\hat{y}(k)\}},
 \end{gather*}  
  where $\hat{y}(k)$ is defined in Proposition \ref{prop Black-Scholes pure q-h} and $k$ is a solution of the equation
\begin{align}\label{warunek na stala generalized q.h.}\nonumber
&C_{BS}(K)-C_{BS}(\hat{y}(k))-(\hat{y}(k)-K)(1-\Phi(e(k)))+C_{BS}(K+c)\\[1ex]
&-\Big[C_{BS}(K+c)-C_{BS}(\hat{y}(k))-(\hat{y}(k)-(K+c))(1-\Phi(e(k)))\Big]\mathbf{1}_{\{K+c\leq\hat{y}(k)\}}=x
\end{align}  
  where
  \begin{gather*}
  e(k):=\frac{\ln(\frac{\hat{y}(k)}{s_0})+\frac{1}{2}\sigma^2T}{\sigma\sqrt{T}}.
  \end{gather*}
In particular, the shortfall of the optimal strategy equals $\{S_T\wedge(K+c)-K\}\mathbf{1}_{\{S_T>\hat{y}(k)\}}$.
 \end{prop}
{\bf Proof:} In view of Proposition \ref{prop o postaci rozwiazan pomocniczych} $(b)$, it follows that 
the solution $\tilde{\varphi}$ of \eqref{problem pomocniczy} has the form
\begin{gather*}
 \tilde{\varphi} =\mathbf{1}_{\{ZH<\frac{1}{k}\}}+\frac{H-L}{H}\mathbf{1}_{\{ZH>\frac{1}{k}\}}.
\end{gather*}
From Theorem \ref{tw o optymalnej strategii (GQH)} we obtain the optimal payoff
$$
\tilde{H}=\tilde{H}(k)=H\tilde{\varphi}=H-L\mathbf{1}_{\{\frac{1}{k}<ZH\}}=(S_T-K)^{+}\mathbf{1}_{\{S_T\leq \hat{y}(k)\}}+(S_T-(K+c))^{+}\mathbf{1}_{\{S_T>\hat{y}(k)\}},
$$ 
which can be also written in the form
\begin{align*}
 \tilde{H}&=
 (S_T-K)^+-(S_T-\hat{y}(k))^+-(\hat{y}(k)-K)\mathbf{1}_{\{S_T>\hat{y}(k)\}}+(S_T-(K+c))^+\\[1ex]
 &-\Big[(S_T-(K+c))^+-(S_T-\hat{y}(k))^+-(\hat{y}(k)-(K+c))\mathbf{1}_{\{S_T>\hat{y}(k)\}}\Big]\mathbf{1}_{\{K+c\leq\hat{y}(k)\}}.
\end{align*}
Finally the constant $k$ is determined by
\begin{align*}
E^Q[\tilde{H}]&=C_{BS}(K)-C_{BS}(\hat{y}(k))-(\hat{y}(k)-K)(1-\Phi(e(k)))+C_{BS}(K+c)\\[1ex]
&-\Big[C_{BS}(K+c)-C_{BS}(\hat{y}(k))-(\hat{y}(k)-(K+c))(1-\Phi(e(k)))\Big]\mathbf{1}_{\{K+c\leq\hat{y}(k)\}},
\end{align*}
which leads to \eqref{warunek na stala generalized q.h.}.
\hfill $\square$

\vskip2mm

\begin{rem} It follows from the proofs of Propositions \ref{prop Black-Scholes pure q-h} and \ref{prop Black-Scholes generalized g-h} that the optimal payoffs for QH and GQH are of the forms
\begin{gather*}
 (S_T-K)^+\mathbf{1}_{\{\frac{1}{k_1}>ZL\}}+(S_T-(K+c))^+\mathbf{1}_{\{\frac{1}{k_1}< ZL\}}, \quad k_1\geq0,\\[1ex]
 (S_T-K)^+\mathbf{1}_{\{\frac{1}{k_2}> ZH\}}+(S_T-(K+c))^+\mathbf{1}_{\{\frac{1}{k_2}< ZH\}}, \quad k_2\geq0,
\end{gather*}
respectively, even for a general form of the shortfall constraint $L$. It follows that, in general, the solutions of QH and GQH 
differ except for the case $L=H$ which corresponds to the case with no shortfall constraint studied in \cite{FL1}. 
\end{rem}

\vskip2mm
\noindent
{\bf Weighted expected shortfall problem (WES)}
\begin{prop}
Let $C_{BS}(K+c)<x<C_{BS}(K)$ and $l(z)=z$. An optimal strategy for a call option $(S_T-K)^+$ with the shortfall constraint $L=c\wedge (S_T-K)^+$ in the Black-Scholes model with parameters satisfying  $\alpha>0$, $\sigma^2>0$ is a replicating strategy for the payoff
 $$
 \tilde{H}=(S_T-K)^+\mathbf{1}_{\{S_T>k\}}+(S_T-(K+c))\mathbf{1}_{\{S_T\leq k\}}
 $$
 with the constant $k\geq 0$ solving
 \begin{align}\label{warunek na stala w (WES)}\nonumber
 C_{BS}(K)&-\left[C_{BS}(K)-C_{BS}(k)-(k-K)(1-\Phi(f(k)))\right]\mathbf{1}_{\{k>K\}}\\[1ex]
 &+\left[C_{BS}(K+c)-C_{BS}(k)-(k-(K+c))(1-\Phi(f(k)))\right]\mathbf{1}_{\{c>K+l\}}
 =x,
 \end{align}
 where
 $$
 f(k):=\frac{\ln(\frac{k}{s_0})+\frac{1}{2}\sigma^2 T}{\sigma\sqrt{T}}.
 $$
 In particular, the shortfall of the optimal strategy equals $\{S_T\wedge(K+c)-S_T\wedge K\}\mathbf{1}_{\{S_T\leq k\}}$.

\end{prop}
{\bf Proof:} By Proposition \ref{prop o postaci rozwiazan pomocniczych} $(c)$ the solution of  \eqref{problem pomocniczy (WES)} is 
$$
\tilde{\varphi}=\mathbf{1}_{\{S_T>k\}}+\frac{H-L}{H}\mathbf{1}_{\{S_T< k\}},
$$
where $k$ is such that the corresponding optimal payoff
\begin{gather*}
 \tilde{H}=\tilde{H}(k)=H\tilde{\varphi}=H-L\mathbf{1}_{\{S_T\leq k\}}=(S_T-K)^+\mathbf{1}_{\{S_T>k\}}+(S_T-(K+c))^+\mathbf{1}_{\{S_T\leq k\}},
\end{gather*}
satisfies $E^Q[\tilde{H}]=x$. The existence and uniqueness of the constant $k$ can be argued as before. 
We obtain its direct characterization by decomposing the optimal payoff $\tilde{H}$ to the combination of the call options with maturities $k,K,K+c$ and applying the Black-Scholes formula.
This leads to \eqref{warunek na stala w (WES)}.
\hfill$\square$

\subsection{Exponential Poisson model}
Let the asset price be given by 
$$
S_t=e^{N_t-\gamma t}, \qquad t\in[0,T],
$$
where $N$ is a Poisson process with intensity $\lambda>0$ under the measure $P$ and $\gamma>0$. The paths of $S$ are neither increasing nor decreasing almost surely, so the model is arbitrage-free, see Theorem 3.2 in \cite{Selivanov} or Proposition 9.9 in \cite{ContTankov}. It is known that any equivalent measure $Q$
is characterized by a new intensity $\lambda_Q(t)=\lambda_Q(\omega,t)\geq0$ of $N$ which means that 
the process
$$
\tilde{N}^Q_t:=N_t-\int_{0}^{t}\lambda_Q(s)ds, \quad t\in[0,T]
$$
is a $Q$-martingale. Using the jump measure language it means that the jump measure of $N$ defined by
$$
\pi(t,A):=\sharp\{s\in[0,t]: \triangle N_s\in A\}, \quad A\subseteq \mathbb{R}, \quad 0\notin\bar{A},
$$
has a compensating measure under $Q$ of the form
$$
\nu_Q(dt,dy):=\lambda_Q(t)\mathbf{1}_{\{y=1\}}(y)dtdy,
$$
that is $\tilde{\pi}_Q(dt,dy):=\pi(dt,dy)-\nu_Q(dt,dy)$ is a compensated measure under $Q$.
The corresponding density of $Q$ with respect to $P$, given by the Girsanov theorem, equals
\begin{gather}\label{exp Poisson gestosc}
\frac{dQ}{dP}=e^{\int_{0}^{T}\ln\left(\frac{\lambda_Q(s)}{\lambda}\right)dN_s-\int_{0}^{T}(\lambda_Q(s)-\lambda)ds}. 
\end{gather}
Let us determine the process $\lambda_Q$ so that $Q$ is a martingale measure. The It\^o formula provides
\begin{align*}
 S_t&=1+\int_{0}^{t}S_{s-}dN_s-\gamma\int_{0}^{t}S_{s-}ds+\sum_{s\in[0,t]}S_{s-}(e^{\triangle N_s}-1-\triangle N_s)\\[1ex]
 &=1-\gamma\int_{0}^{t}S_{s-}ds+\sum_{s\in[0,t]}S_{s-}(e^{\triangle N_s}-1)\\[1ex]
 &=1-\gamma\int_{0}^{t}S_{s-}ds+\int_{0}^{t}\int_{\mathbb{R}}S_{s-}(e^{y}-1)\tilde{\pi}_Q(ds,dy)+\int_{0}^{t}\int_{\mathbb{R}}S_{s-}(e^{y}-1)\nu_Q(ds,dy)\\[1ex]
 &=1+\int_{0}^{t}\int_{\mathbb{R}}S_{s-}(e^{y}-1)\tilde{\pi}_Q(ds,dy)+\int_{0}^{t}S_{s-}\Big((e-1)\lambda_Q(s)-\gamma\Big)ds.
\end{align*}
It follows that $S$ is a local martingale under $Q$ if and only if
\begin{gather}\label{lambda Q}
\lambda_Q(t)\equiv\lambda_Q=\frac{\gamma}{e-1}.
\end{gather}
Hence from \eqref{exp Poisson gestosc} and \eqref{lambda Q} it follows that the model admits only one martingale measure 
$Q$ with the density of the form
$$
\frac{dQ}{dP}=Z=e^{\ln\big(\frac{\lambda_Q}{\lambda}\big)N_T-(\lambda_Q-\lambda)T}=C \cdot S_T^{\ln\big(\frac{\gamma}{\lambda(e-1)}\big)},
$$
where
$$
C:=e^{T\big(\lambda-\frac{\gamma}{e-1}-\gamma\ln\big(\frac{\lambda(e-1)}{\gamma}\big)\big)}.
$$
It follows that the price of a call option $(S_T-K)^+, K\geq0$ is equal to
$$
C_{EP}(K):=E^Q[(e^{N_T-\gamma T}-K)^+]=\sum_{n=\lceil\ln K+\gamma T\rceil}^{+\infty}(e^{-\gamma T}-K)e^{-\frac{\gamma}{e-1}T}\frac{(\frac{\gamma}{e-1}T)^n}{n!},
$$
where $\lceil a\rceil:=\inf\{n\in\mathbb{N}: n\geq a\}$.

In this model the quantile hedging problem does not have a solution because the set $\tilde{A}$ described by Proposition \ref{prop pure quantile hedging} may not exist. Since the law of $ZL$ is not atom-free, Proposition \ref{prop o postaci rozwiazan pomocniczych} can not be applied. Below we solve the generalized quantile hedging problem by direct application of Lemma \ref{lemat NP generalized} in the case when $H$ is
a call option and  the coefficients satisfy $1<\frac{\lambda(e-1)}{e}<\gamma$. Other cases can be treated in a similar manner.
Notice that for $x=C_{EP}(K+c)$ the replicating strategy for $(S_T-(K+c))^+$ is optimal.

\begin{prop}
 Let $H=(S_T-K)^+$ and $L=c\wedge (S_T-K)^+$ with $c,K\geq 0$. If 
 \begin{gather}\label{exp poisson wspolczynniki}
1<\frac{\lambda(e-1)}{e}<\gamma,  
 \end{gather}
 then an optimal solution to the 
 generalized quantile hedging problem with initial capital $x$ satisfying
 $$
C_{EP}(K+c)< x<C_{EP}(K)
 $$
is a replicating strategy for the payoff 
 \begin{align}\label{exp Poisson - optimal payoff}
 \tilde{H}&=(S_T-K)^{+}\mathbf{1}_{\{S_T<\hat{y}(k)\}}+(S_T-(K+c))^{+}\mathbf{1}_{\{S_T\geq\hat{y}(k)\}}+\gamma\Big(c\wedge (S_T-K)^+\Big)\mathbf{1}_{\{S_T=\hat{y}(k)\}}.
\end{align}
Here, $\hat{y}(k)$ stands for the unique solution of the equation
 $$
 y-K=\frac{1}{Ck}y^{\ln\big(\frac{\lambda(e-1)}{\gamma}\big)}; \quad y\geq0,
 $$
the constant $k$ is determined as 
 \begin{align}\nonumber\label{k exp poisson}
 k:=\inf\Big\{ &u\geq0: f(u):=C_{EP}(K)-C_{EP}(\hat{y}(u))-(\hat{y}(u)-K)Q(S_T>\hat{y}(u))+C_{EP}(K+c)\\[1ex]
 &-\big[C_{EP}(K+c)-C_{EP}(\hat{y}(u))-(\hat{y}(u)-(K+c))Q(S_T\geq \hat{y}(u))\big]\mathbf{1}_{\{K+c\leq \hat{y}(u)\}}\leq x\Big\},
 \end{align}
 and $\gamma$ solves the equation
 \begin{gather}\label{gamma exp poisson}
 f(k)+\gamma\Big(c\wedge(\hat{y}(k)-K)\Big)Q(S_T=\hat{y}(k))=x.
 \end{gather}
 The corresponding shortfall equals
 $$
 \Big(c\wedge (S_T-K)^+\Big)\mathbf{1}_{\{S_T>\hat{y}(k)\}}+(1-\gamma)\Big(c\wedge (\hat{y}(k)-K)^+\Big)\mathbf{1}_{\{S_T=\hat{y}(k)\}}.
 $$
 \end{prop}
 \noindent
 {\bf Proof:} Solving \eqref{problem pomocniczy} with the use of  Lemma \ref{lemat NP generalized} yields the optimal payoff
 \begin{align*}
 \tilde{H}=H\tilde{\varphi}&=H \mathbf{1}_{\{\frac{dP}{d\tilde{Q}}>a\}}+(H-L+\gamma L)\mathbf{1}_{\{\frac{dP}{d\tilde{Q}}=a\}}
 +(H-L)\mathbf{1}_{\{\frac{dP}{d\tilde{Q}}<a\}}\\[1ex]
 &=H-L\mathbf{1}_{\{\frac{dP}{d\tilde{Q}}\leq a\}}+\gamma L \mathbf{1}_{\{\frac{dP}{d\tilde{Q}}=a\}}\\[1ex]
 &=(S_T-K)^+-\Big(c\wedge(S_T-K)^+\Big)\mathbf{1}_{\{\frac{1}{k}\leq Z(S_T-K)^+\}}+\gamma \Big(c\wedge(S_T-K)^+\Big)\mathbf{1}_{\{\frac{1}{k}= Z(S_T-K)^+\}}
 \end{align*}
 where $\frac{d\tilde{Q}}{dQ}=\frac{H}{E^Q[H]}$ and the constants $a,k\geq 0,\gamma\in[0,1]$ should be such that $E^Q[\tilde{H}]=x$. Using \eqref{exp poisson wspolczynniki} we obtain the alternative characterization
 $$
 \tilde{H}=\tilde{H}(k,\gamma)=(S_T-K)^+-\Big(c\wedge(S_T-K)^+\Big)\mathbf{1}_{\{S_T\geq\hat{y}(k)\}}+\gamma \Big(c\wedge(S_T-K)^+\Big)\mathbf{1}_{\{S_T=\hat{y}(k)\}}.
 $$
 To characterize $k$ and $\gamma$ let us notice that the function
 $$
 f(z):=E^Q\Big[(S_T-K)^+-\Big(c\wedge(S_T-K)^+\Big)\mathbf{1}_{\{S_T\geq\hat{y}(z)\}}\Big]; \quad z\geq0,
 $$
 is decreasing, c\`adl\`ag and satisfies
 \begin{align}\label{EP f pierwszy warunek}
  \lim_{z\rightarrow 0}f(z)=C_{EP}(K), \lim_{z\rightarrow +\infty}f(z)=C_{EP}(K+c), \\[1ex]\label{EP f drugi warunek}
  \triangle f(z)=-\Big(c\wedge (\hat{y}(z)-(K+c))^+\Big)Q(S_T=\hat{y}(z)).
 \end{align}
 Since $C_{EP}(K)\leq x<C_{EP}(K+c)$ and \eqref{EP f pierwszy warunek} holds, the constant $
 k:=\inf\{z\geq 0: f(z)\leq x\}$
 is well defined. Moreover, there exists $\gamma\in[0,1]$ such that  $x-f(k)=\gamma (-\triangle f(k))$.
 In view of \eqref{EP f drugi warunek} this yields
 $$
 x=f(k)+\gamma \Big(c\wedge (\hat{y}(k)-(K+c))^+\Big)Q(S_T=\hat{y}(k)),
 $$
 which means that $E^Q[\tilde{H}(k,\gamma)]=x$ as required. To obtain \eqref{k exp poisson} and
 \eqref{gamma exp poisson} one decomposes $\tilde{H}$ into the form
   \begin{align*}
 \tilde{H}&=
 (S_T-K)^+-(S_T-\hat{y}(k))^+-(\hat{y}(k)-K)\mathbf{1}_{\{S_T>\hat{y}(k)\}}+(S_T-(K+c))^+\\[1ex]
 &-\Big[(S_T-(K+c))^+-(S_T-\hat{y}(k))^+-(\hat{y}(k)-(K+c))\mathbf{1}_{\{S_T\geq\hat{y}(k)\}}\Big]\mathbf{1}_{\{K+c\leq\hat{y}(k)\}}\\[1ex]
 &+\gamma\Big( c\wedge(S_T-K)^+\Big)\mathbf{1}_{\{S_T=\hat{y}(k)\}}.
\end{align*}
and notices that $E^Q[\tilde{H}]=f(k)+\gamma\Big( c\wedge(\hat{y}(k)-K)\Big)Q(S_T=\hat{y}(k))$. 
\hfill$\square$

\end{document}